\documentclass[useAMS,usenatbib,twocolumn]{mn2e}
\usepackage{graphics,epsfig}
\usepackage{graphicx,times}
\usepackage[usenames,dvipsnames]{color}
\usepackage{amsmath,amssymb,bm,url}
\usepackage{etoolbox}
\def \ff {{\bm f}}

\newcommand{\DD}{{\rm D} {}}
\newcommand{\OO}{\bm{\Omega}}
\newcommand{\zzz}{\hat{\bm{z}}}
\newcommand{\Tab}[1]{Table~\ref{#1}}
\newcommand{\mean} [1] {\overline #1}

\def\MA{M_{\rm A}}
\def\tautd{\tau_{\rm td}}

\def\zstar{z_\star}
\def\xc{x_{\rm c}}
\def\yc{y_{\rm c}}
\def\zc{z_{\rm c}}

\def \del2z {\partial^{2}_{z}}

\def \Beq {B_{\rm eq}}

\def \AAA {{\bm A}}

\def \UU {{\bm U}}

\def \JJ {{\bm J}}
\newcommand{\EE}{\bm{{E}}}
\newcommand{\SSSS}{\mbox{\boldmath ${\sf S}$} {}}

\def \grav {{\bm g}}

\newcommand{\nab}{\mbox{\boldmath $\nabla$} {}}

\newcommand{\etal}{et al.}

\newcommand{\Eq}[1]{Eq.~(\ref{#1})}

\newcommand{\Eqss}[2]{Eqs.~(\ref{#1})--(\ref{#2})}

\newcommand{\bra}[1]{\left\langle #1\right\rangle}

{}
\newcommand{\dd}{{\rm d} {}}
\newcommand{\Co}{{\rm Co}}
\newcommand{\G}{\,{\rm G}}
\def \Rm  {\mbox{Re}_{\rm M}}

\def \Rey  {\mbox{Re}}

\def \kf  {k_{\rm f}}

\def \urms  {u_{\rm rms}}

\def \BB {\bm B}

\def \etat {\eta_{\rm t}}
\def\etatz{\eta_{\rm t0}}

\def \Rm  {\mbox{Re}_{\rm M}}
\def \Pm  {\mbox{Pr}_{\rm M}}
\def \cs  {c_{\rm s}}
\def \kf  {k_{\rm f}}

\def \urms {u_{\rm rms}}

\def \BB {\bm B}

\def\onethird{{\textstyle{1\over3}}}

\def \half {{\textstyle{1\over2}}}

\newcommand{\Fig}[1]{Fig.~\ref{#1}}
\newcommand{\FFig}[1]{Figure~\ref{#1}}
\newcommand{\Figs}[2]{Figs.~\ref{#1} and \ref{#2}}

\def\drawing #1 #2 #3 {
\begin{center}
\setlength{\unitlength}{1mm}
\begin{picture}(#1,#2)(0,0)
\put(0,0){\framebox(#1,#2){#3}}
\end{picture}
\end{center} }

\graphicspath{{./fig/}{./png/}}
\def\blue{\textcolor{black}}
\title[Turbulent reconnection of magnetic bipoles]
{Turbulent reconnection of magnetic bipoles in stratified turbulence}
\author[S. Jabbari et al.]{
S. Jabbari$^{1,2}$
\thanks{E-mail: sarahjab@kth.se},
A. Brandenburg$^{1,2,3,4}$,
Dhrubaditya Mitra$^{1}$,
N. Kleeorin $^{5,1}$,
I. Rogachevskii $^{5,1}$
\\
$^1$Nordita, KTH Royal Institute of Technology and Stockholm University,
    Roslagstullsbacken 23, SE-10691 Stockholm, Sweden\\
$^2$Department of Astronomy, AlbaNova University Center,
    Stockholm University, SE-10691 Stockholm, Sweden\\
$^3$JILA and Department of Astrophysical and Planetary Sciences,
    Box 440, University of Colorado, Boulder, CO 80303, USA\\
$^4$Laboratory for Atmospheric and Space Physics,
    3665 Discovery Drive, Boulder, CO 80303, USA\\
$^5$Department of Mechanical Engineering, Ben-Gurion University of the Negev,
    POB 653, Beer-Sheva 84105, Israel\\
}
\begin{document}
\date{\today,~ $ $Revision: 1.240 $ $}
\pagerange{\pageref{firstpage}--\pageref{lastpage}} \pubyear{}

\maketitle

\label{firstpage}

\begin{abstract}
We consider strongly stratified forced turbulence
in a plane-parallel layer
with helicity and corresponding large-scale dynamo
action in the lower part and non-helical turbulence in the upper.
The magnetic field is found to develop strongly concentrated bipolar
structures near the surface.
They form elongated bands with a sharp interface between opposite polarities.
Unlike earlier experiments with imposed magnetic field, the inclusion
of rotation does not strongly suppress the formation of these structures.
We perform a systematic numerical study of this phenomenon by varying
magnetic Reynolds number, scale separation ratio, and Coriolis number.
We focus on the formation
of a current sheet between bipolar regions
where reconnection of oppositely oriented field lines occurs.
We determine the reconnection rate by measuring either
the inflow velocity in the vicinity of the current sheet
or by measuring the electric field in the reconnection region.
We demonstrate that for large Lundquist numbers, $S>10^3$,
the reconnection rate is nearly independent of $S$
in agreement with results of recent numerical simulations
performed by other groups in simpler settings.
\end{abstract}

\section{Introduction}

The mechanism for the formation of sunspots
and active regions is still not understood.
One popular model assumes that the solar dynamo generates thin, strong
magnetic flux tubes of $\sim 10^5\G$ near the tachocline \citep{DSC93}.
Part of these tubes can become magnetically buoyant and rise to the surface
creating sunspots and active regions \citep{P55,CSD95}; see also
\cite{Fan09} for a review.
So far neither numerical \citep{GK11} nor
observational \citep{Fan09,ZBK13,Get15} studies have confirmed this scenario.
Furthermore, the flux tubes are expected to expand as they rise,
hence their strength weakens and some sort of re-amplification mechanism
must complement this model to match the observational properties of
sunspots.

An alternative mechanism for the formation of active regions and
sunspots is based on the negative effective (mean-field) magnetic
pressure instability (NEMPI).
Analytical studies \citep{KRR89,KRR90,KMR93,KMR96,KR94,RK07} supported by
direct numerical simulations \citep{BKKR12,KBKMR12,KBKKR16} have shown that in
stratified turbulence in the presence of a background magnetic field
the effective magnetic pressure
(the sum of turbulent and non-turbulent contributions)
can become negative.
This effect can give rise to a large-scale
instability \citep{RK07,KBKMR13}, i.e., NEMPI, which can lead to the
concentration of a weak background magnetic field.
Once the field becomes strong enough -- more than the local equipartition value
-- the effective magnetic pressure is no longer negative and NEMPI
is not excited.
Direct numerical simulations (DNS) of stratified turbulence have demonstrated
that NEMPI can produce
magnetic field concentrations \citep{BKR10,BKKMR11,KBKMR13,Jabbari14}
from background magnetic fields that can be either perpendicular or parallel
to the density gradient in both spherical and rectangular domains.
In the latter case, spot-like structures near the surface
\citep{BKR13,BGJKR14} are obtained.
A further generalization to a two-layer model \citep{WLBKR13,WLBKR16}
with non-helical forcing in the lower layer and no forcing in the upper
has been successful in generating bipolar magnetic structures with intriguing
dynamical behavior.
Furthermore, mean-field simulations have shown that the concentration of a
background magnetic field by NEMPI can operate even in the presence of
dynamo action \citep{Jabbari13}.

Nevertheless, there are two limitations to the aforementioned
studies: (a) it is necessary to have a weak
initial background magnetic field to excite NEMPI,
and (b) the maximum strength of the magnetic field
in the nonlinear stage of NEMPI can be at most three times larger than
the local equipartition value.
\cite{Mit14} and \cite{Jabbari15} have circumvented these limitations
respectively in their Cartesian and spherical models
by using a two-layer arrangement of forced stratified
turbulence in which the bottom layer is helically forced and the top
layer is non-helically forced.
In both of these cases, a large-scale dynamo develops in
the bottom layer and provides a background magnetic field that is
concentrated by stratified turbulence in the top layer to generate
intense magnetic structures of strengths that can be close to five
times the equipartition value.
It is not clear that NEMPI is the relevant mechanism that gives rise
to the magnetic structures observed by \cite{Mit14} or \cite{Jabbari15}.
Although \cite{Mit14} have not measured the effective magnetic pressure,
they did detect large-scale downflows at the location
of the magnetic flux concentrations.
Similar downflows have been found previously
in forced turbulence with an imposed vertical field \citep{BKR13,BGJKR14},
where NEMPI was found to lead to magnetic spot formation.
In the work of \cite{Jabbari15}, NEMPI could only be excited
in those parts of the domain where the strength of the dynamo-generated
field was sufficiently below the equipartition magnetic field.
Nevertheless, also in that case formation of spots coincided with downflows.

The purpose of the present study is two-fold.
On the one hand, using the model of \cite{Mit14}, we
perform a systematic numerical study of the formation
and decay of bipolar regions by varying different parameters
of the problem, in particular the magnetic Reynolds number
and the scale separation ratio.
Furthermore, we study the effects of rotation through the Coriolis term
in the same model.
As emphasized by \cite{Mit14} and \cite{Jabbari15}, the lifetime of the
sharp interface between the bipolar regions is much longer than
what one estimates from the effects of turbulent diffusion.
This suggests that the sharp interface is constantly being maintained by
converging flows, which lead to the formation of a current sheet between
two polarities and the occurrence of turbulent reconnection.

Magnetic reconnection is a fundamental plasma process
that is believed to play an important role in different
astrophysical, geophysical, and laboratory plasma phenomena,
e.g., solar flares, coronal mass ejections, coronal heating,
magnetospheric substorms and tearing mode instabilities
in magnetic confinement fusion devices \citep{PR14,ZWY09,LSU13}.
A classical model of reconnection was suggested by \cite{P57} and \cite{S58};
see also their later works \citep{S69,P94}.
According to the Sweet-Parker model, the reconnection rate is proportional
to the square root of the magnetic diffusivity of the plasma.
This would imply that in the astrophysically relevant limit of
very small magnetic diffusivity (or very large Lundquist number)
the Sweet-Parker reconnection rate would go to zero.
Hence, for reconnection to be relevant in the astrophysical context
it is necessary to find models of fast reconnection in which the
reconnection rate is independent of Lundquist number in the
asymptotic limit of large Lundquist number.
Recently, fast reconnection has been studied
in DNS of turbulent magnetohydrodynamics (MHD) in both two and
three dimensions \citep{LUS09,KLV09,HB10,LSS12,B13,Oishi15},
and at least two competing models:
by \cite{LV99,ELV11} and by \cite{ULS10,LSU13}
have been proposed, see, e.g., \cite{LEVK15} for a review.
To investigate the role of magnetic reconnection in our model, we
zoom in on the flow around the sharp interface,
study the dynamics of the current sheet in this region
and measure the reconnection rate
to determine which regime of turbulent reconnection
is relevant to our system.

\section{The model}
\label{DNSmodel}

\subsection{Basic equations}

To perform DNS of an isothermally stratified layer, we solve the equations
for the velocity $\UU$, the magnetic vector potential $\AAA$, and
the density $\rho$ and, in some cases, also in the presence of nonvanishing
angular velocity $\OO=\Omega\zzz$,
\begin{equation}
\rho{\frac{\DD\UU}{\DD t}}=\JJ\times\BB
-2\OO\times\rho\UU-\cs^2\nab\rho+\nab\cdot(2\nu\rho\SSSS)
+\rho(\ff+\grav),
\label{DUDt}
\end{equation}
\begin{equation}
{\partial\AAA\over\partial t}=\UU\times\BB+\eta\nabla^2\AAA,
\label{DADt}
\end{equation}
\begin{equation}
{\partial\rho\over\partial t}=-\nab\cdot\rho\UU,
\label{drhodt}
\end{equation}
where the operator $\DD/\DD t=\partial/\partial t+\UU\cdot\nab$ is
the advective derivative,
${\sf S}_{ij}=\half(U_{i,j}+U_{j,i})-\onethird\delta_{ij}\nab\cdot\UU$
is the traceless rate of strain tensor (the commas denote
partial differentiation), $\nu$ is the kinematic viscosity,
$\cs$ is the isothermal sound speed, $\mu_0$ is the vacuum permeability,
$\eta$ is the magnetic diffusivity,
$\BB=\nabla\times\AAA$ is the magnetic field,
and $\JJ=\nabla\times\BB/\mu_0$ is the current density.

We perform simulations in a cubic domain of size $L^3$.
This implies that the smallest wavenumber which fits into the box is 1
($k_{1}=2\pi/L=1$).
We apply the same boundary condition as \cite{Mit14}, i.e., we use
periodic boundary conditions in the $x$ and $y$ directions,
stress-free perfect conductor boundary conditions at the bottom of the domain,
($z=-L/2$) and stress-free vertical field conditions at the top ($z=+L/2$).

The stratification is isothermal with constant gravity given by
$\bm{g}=(0,0,-g)$, so the density scale height is $H_{\rho}=c_{s}^2/g$.
In all the cases considered below we have $k_{1}H_{\rho}=1$
and $L/H_{\rho}=2\pi$.
In this setup the density contrast across the domain is
$\exp(L_z/H_{\rho})=\exp 2\pi\approx535$.
Since we have adopted an isothermal equation of state, there is
no possibility of convection.
We apply random volume forcing to drive turbulence.
It is defined by a function $f$ that is $\delta$-correlated in time and
monochromatic in space.
It consists of random non-polarized waves whose direction and phase
change randomly at each time step.
To simulate the two-layer model of \cite{Mit14},
we define the forcing profile such that we have helical forcing
in the lower part of the domain ($z<\zstar$) and non-helical forcing
in the upper ($z>\zstar$).
Here, $\zstar$ is the position of the border between helical
and non-helical forcing; in our model we choose $\zstar=-H_{\rho}$.
The helical forcing leads to the generation of a large-scale magnetic field
in the lower layer due to $\alpha^2$ dynamo action.
The field then diffuses to the upper layer where the magnetic bipolar spots
are expected to form.
For more details regarding the forcing profile, see \cite{Mit14}.

This setup is chosen to demonstrate the physical
effects in isolation.
In particular, the region of the dynamo generating large-scale weakly
non-uniform magnetic field is separated from the region where
the strongly non-uniform bipolar magnetic region is formed.
This arrangement can also mimic a nonuniform spatial distribution
of kinetic helicity and $\alpha$ effect in the solar convective zone,
e.g., the $\alpha$ is larger in the deeper parts of the convective zone
\citep{Kri84}.

\subsection{Parameters of the simulations}

To solve \Eqss{DUDt}{drhodt}, we perform DNS with
the {\sc Pencil Code}\footnote{\url{https://github.com/pencil-code}}.
It uses sixth-order explicit finite differences in space
and a third-order accurate time-stepping method.
We use a numerical resolution of $256\times252\times256$ mesh points
in the $x$, $y$, and $z$ directions.
We choose our units such that $\cs=g=\mu_0=1$.
Our simulations are characterized by the
fluid Reynolds number, $\Rey\equiv\urms/\nu\kf$,
and the magnetic Prandtl number, $\Pm=\nu/\eta$, so
the magnetic Reynolds number is $\Rm\equiv \Rey \, \Pm = \urms/\eta\kf$.
Here, $\kf/k_1$ is the forcing wavenumber, which takes a value of $30$ in
most of our simulations.
We also study the case $\kf/k_1=5$.
In all runs we keep, $\Pm=0.5$ and vary $\Rm$.

As the value of the turbulent velocity is set by the local strength
of the forcing, which is uniform, the turbulent velocity is also
statistically uniform over depth, and therefore we choose to define
$\urms$ as the root-mean-square velocity based on a volume average
in the statistically steady state.
On the other hand, the density varies over several orders of magnitude as
a function of depth and hence we define the mean density $\mean\rho$ as a
a horizontally and temporally averaged density at each depth.
The magnetic field is expressed in units of the local equipartition value,
$\Beq\equiv\sqrt{\mu_0\mean\rho} \, \urms$.
Time is measured in turbulent-diffusive times,
$\tautd=(\etatz k_1^2)^{-1}$, where $\etatz=\urms/3\kf$ is
the estimated turbulent magnetic diffusivity.

Most of the parameters are similar to those of \cite{Mit14}, but we vary
$\Rm$ and also consider cases with rotation.
We also study the effect of the scale separation ratio, $\kf/k_1$, by changing
$\kf$ to investigate its effect on the formation and evolution of
bipolar structures.
\Tab{Tab1} shows all runs with their parameters.

We characterize the reconnection of the bipolar magnetic structures
by the Lundquist number, $S$:
\begin{equation}
S=V_{\rm A} {\cal L} /\eta ,
\end{equation}
where $V_{\rm A}=B/\sqrt{\mu_{0}\rho}$ is the Alfv\'en velocity,
and ${\cal L}$ is a typical length scale, which is here taken to be
the length of the current sheet.

\begin{table}\caption{
Summary of the runs.
The reference run is shown in bold.
}\vspace{12pt}\centerline{\begin{tabular}{lrlcccccl}
Run &$\Rm$&$\kf/k_1$&$\zstar/H_{\rho}$& $\lambda/\etatz k_1^2$ \\
\hline
D     &  16  & 30 &$-1$& 0.042\\ 
RM0   &  10  &  30 & $-1$& $\!$0.05\\
{\bf RM1}& {\bf 50} & {\bf 30} &${\bf -1}$& {\bf 0.014}\\
RM1zs   & 50 &  30 & $\pi$& 0.013 \\
RM1k   &  50 & 5 & $-1$& 0.024 \\
RM2   &  130 &  30 &$-1$& 0.005 \\
RM3   &  260 &  30 &$-1$& 0.002 \\
RM4   &  300 &  30 &$-1$& 0.001  \\
\label{Tab1}\end{tabular}}
\end{table}

\begin{figure}\begin{center}
\includegraphics[width=.95\columnwidth]{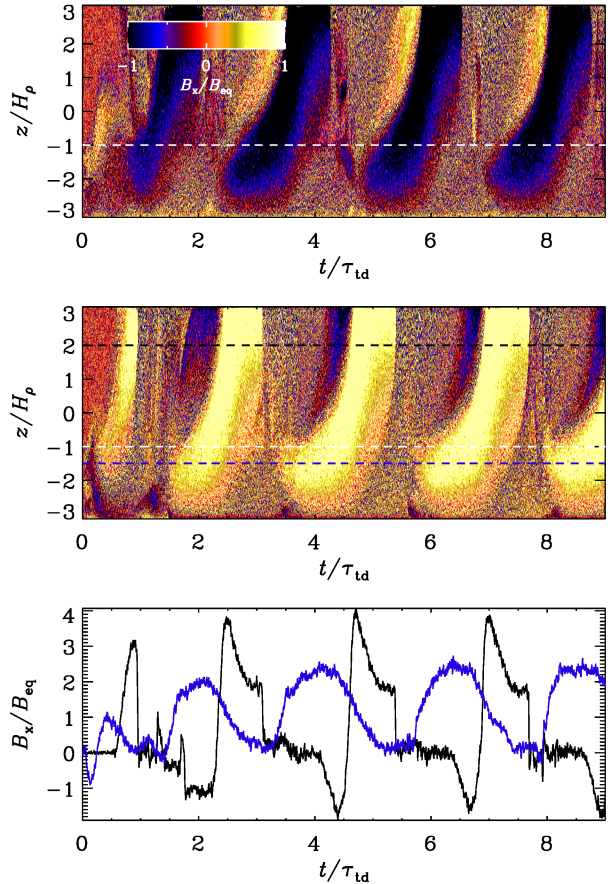}
\end{center}\caption[]{
Butterfly diagrams $B(\xc,\yc,z,t)/\Beq(z)$ for Run~RM1
through cross-sections $\xc/H_\rho=\pi$, and either
$\yc/H_\rho=1.8$ (top panel) or $-1.8$ (middle panel),
as well as $B_x(\xc,\yc,\zc,t)/\Beq$ through $\xc/H_\rho=\pi$,
$\yc/H_\rho=-1.8$, for both $\zc/H_\rho=2$ (black line) and
$\zc/H_\rho=-1.5$ (blue line) versus time (lower panel).
The white dashed lines in the upper two panels indicate the position of
$\zstar$ and in the second panel the dashed blue and black horizontal lines
show the locations where $B_x(\xc,\yc,\zc,t)/\Beq$ is plotted vs.\ $t$.
}\label{pbutt50}\end{figure}

\begin{figure}\begin{center}
\includegraphics[width=\columnwidth]{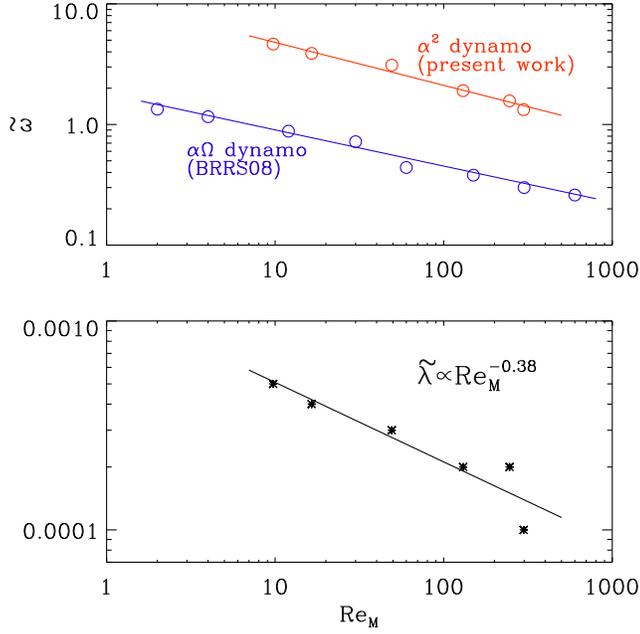}
\end{center}\caption[]{
Upper panel: normalized dynamo frequency, $\tilde{\omega}$, as a function of $\Rm$.
The solid lines show the best fit to our data points (red filled circles) and
the data points of \cite{BRRS08} (blue open circles).
Lower panel: normalized growth rate of dynamo, $\tilde{\lambda}$, as a function of $\Rm$.
The solid line shows the best fit to our data points.
}\label{fft}\end{figure}

\section{Properties of the dynamo}

The magnetic field in our model is the result of a large-scale dynamo.
We recall that the forcing in the momentum equation is fully helical
in the lower $30\%$ of the box and non-helical in the rest.
We expect that the $\alpha^2$ dynamo generates an exponentially growing
magnetic field during the early phase when the field is weak.
The dynamo-generated field has a periodic behavior
with dynamo waves propagating in the domain, where the forcing is helical
($z/H_\rho<-1$).
To study the dynamo properties, we plot the butterfly diagram in \Fig{pbutt50}.
The upper panel shows the butterfly diagram at $y/H_\rho=1.8$ and
the middle panel presents the same plot for $y/H_\rho=-1.8$.
The speed of upward motion increases as one moves
toward the surface for both polarities.
The nondimensional growth rate is given as $\tilde{\lambda}=\lambda/\urms\kf$,
where $\lambda=\dd \ln B_{\rm rms}/\dd t$, and its value
decreases with increasing magnetic Reynolds number (see \Tab{Tab1}
and the lower panel of \Fig{fft}).
The large-scale magnetic field expands upward into the region
with non-helical forcing due to turbulent magnetic diffusion.
In this region, density is lower, so the field strongly exceeds
the equipartition value.
Here the field evolution is highly nonlinear and driven by the dynamo wave
from beneath.

To measure the period of the dynamo cycle, we plot
in the lower panel of \Fig{pbutt50} the value of $B_{x}/\Beq$
as a function of time for two different depths,
$z/H_\rho=-1.5$ (inside the helical region; blue curve) and
$z/H_\rho=2$ (near the surface; black curve).
In Run~D the value of the period is about $1.6\tautd$ for $\Rm=16$.
This is consistent with the result of \cite{Mit14},
where a period of the dynamo wave of $1.5\tautd$ was determined.

In the upper panel of \Fig{fft} we show the dependence of the
normalized dynamo frequency, $\tilde{\omega}=\omega/\etatz k_1^2$,
on the magnetic Reynolds number, $\Rm$.
The fit overplotted on the data has the form
\begin{equation}
\tilde{\omega}\approx 11\,\Rm^{-0.36}.
\end{equation}
Although the large-scale dynamo that develops in this
problem is an $\alpha^2$ dynamo,
the functional dependence of $\omega$ on $\Rm$ is similar to that
of a nonlinear
$\alpha\Omega$ dynamo in a Cartesian domain with linear shear \citep{KB09}.
In that case, $\omega\propto\etat k_1^2$ is proportional to the quenched
turbulent magnetic diffusivity, $\etat$.
In a separate study of a nonlinear $\alpha^2$ dynamo, $\etat$ is found to be
proportional to $\Rm^{-0.3}$ for $2\leq\Rm\leq600$ \citep{BRRS08}.
The prefactor is, however, about 6 times larger in the present case of
an $\alpha^2$ dynamo compared to the earlier $\alpha\Omega$ dynamos,
where $\tilde\omega\approx1.8\,\Rm^{-0.3}$ has been found; cf.\ \Fig{fft}.
However, for much larger magnetic Reynolds numbers, $\tilde\omega$
as well as the growth rate of the large-scale dynamo instability
might become independent of $\Rm$.
Interestingly, the normalized growth of the dynamo displays a similar
dependence on $\Rm$; see the lower panel of \Fig{fft}.
Specifically, we find $\tilde{\lambda}\approx 0.0012\,\Rm^{-0.38}$.

\begin{figure}\begin{center}
\includegraphics[width=.99\columnwidth]{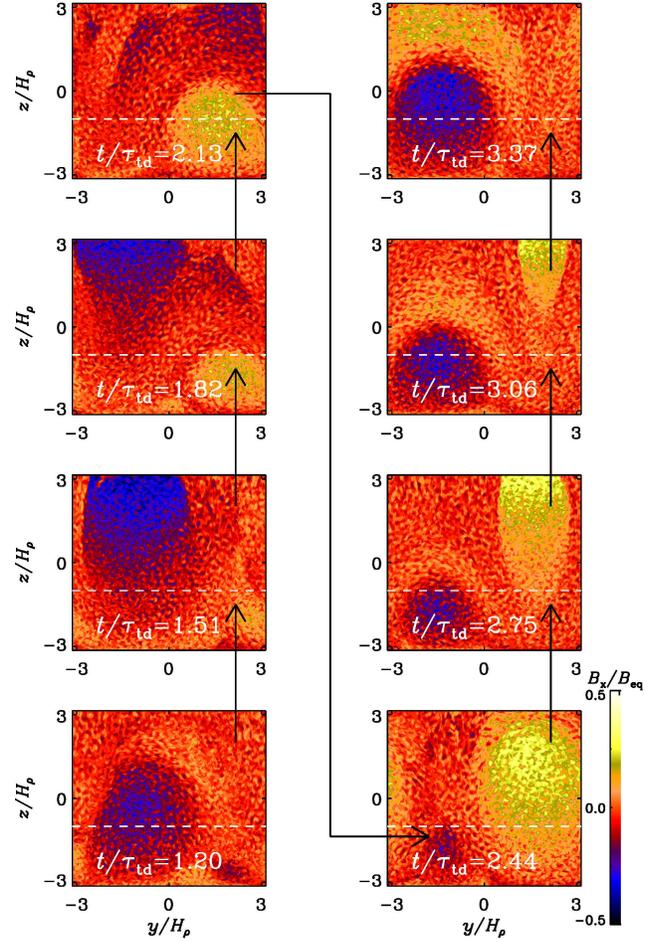}
\end{center}\caption[]{
Time evolution of $B_{x}/\Beq$ in the $yz$ plane
through $x/H_\rho=\pi$ for Run~RM1.
}\label{pbx_yz}\end{figure}

The cyclic behavior of the dynamo-generated magnetic field is also shown in a
series of snapshots in \Fig{pbx_yz}, where we present
$\mean B_{x}/\Beq$ in the $yz$ plane at different times.
As one can see, dynamo waves propagate toward the surface and
the evolution of the polarities is similar.

\section{Magnetic structures}

As mentioned above, the evolution and formation of the magnetic structures
is similar to that of \cite{Mit14} and \cite{Jabbari15}.
For $\Rm=$10, 16, and 50, bipolar magnetic structures of super-equipartition
strength form in about half a turbulent-diffusive time and continue to evolve.
For higher $\Rm$, structures form at later times and survive much longer
compared to the case with a smaller $\Rm$ of 10 or 16.
However, the type of structures are otherwise similar for all $\Rm$.

\begin{figure}\begin{center}
\includegraphics[width=.9\columnwidth]{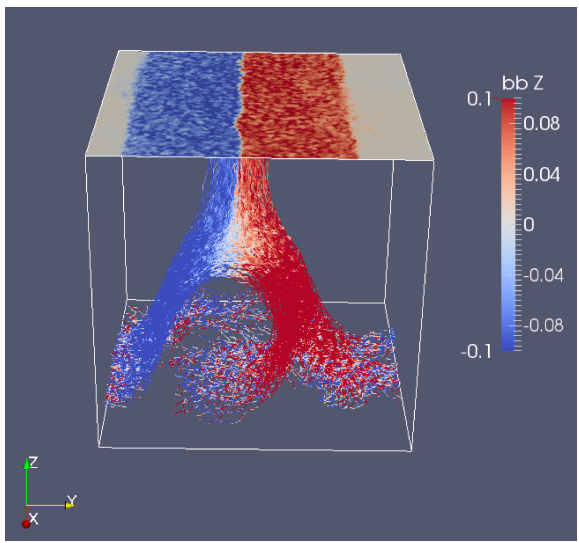}
\end{center}\caption[]{
Three-dimensional visualization of vertical magnetic field, $B_{z}$ at the
surface (color-coded) together with \blue{three-dimensional volume rendering of
the vertical component of} the magnetic field for Run~RM1.
}\label{bz3D}\end{figure}

\begin{figure}\begin{center}
\includegraphics[width=.99\columnwidth]{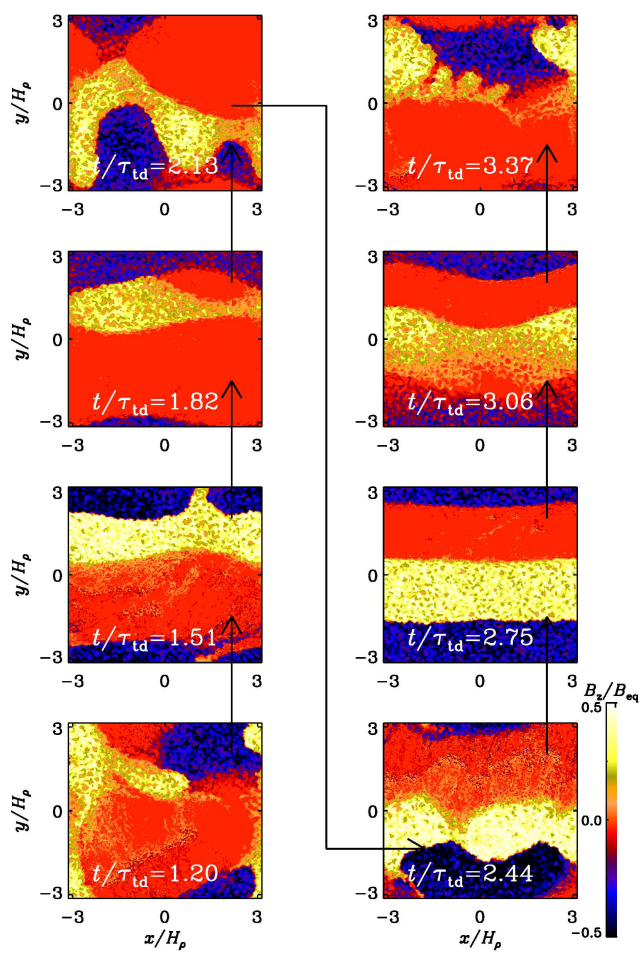}
\end{center}\caption[]{
Time evolution of $B_z/\Beq$ in the $xy$ plane
at the top surface for Run~RM1.
}\label{pbz_xy}\end{figure}

\subsection{Production of sharp fronts}

In this paper we are interested in the time when structures develop
and form ``stripy'' patterns at the surface (see \Figs{bz3D}{pbz_xy}).
We concentrate on the phase in the evolution when
different polarities move close together
to form a current sheet between magnetic fields of the opposite polarities.
\FFig{pbz_xy} illustrates the time evolution of $B_{z}/\Beq$
at the surface ($z/H_\rho=\pi$).
One can see the formation of a sharp boundary between two polarities
in the right column, third panel of this figure ($t/\tautd=2.75$).
It is clear from \Fig{pbz_xy} that the characteristic time of the formation
of the elongated structures is of the order of the period of the dynamo wave,
i.e., the turbulent-diffusion time
(compare this figure with \Fig{pbx_yz} for instance).

\begin{figure}\begin{center}
\includegraphics[width=.99\columnwidth]{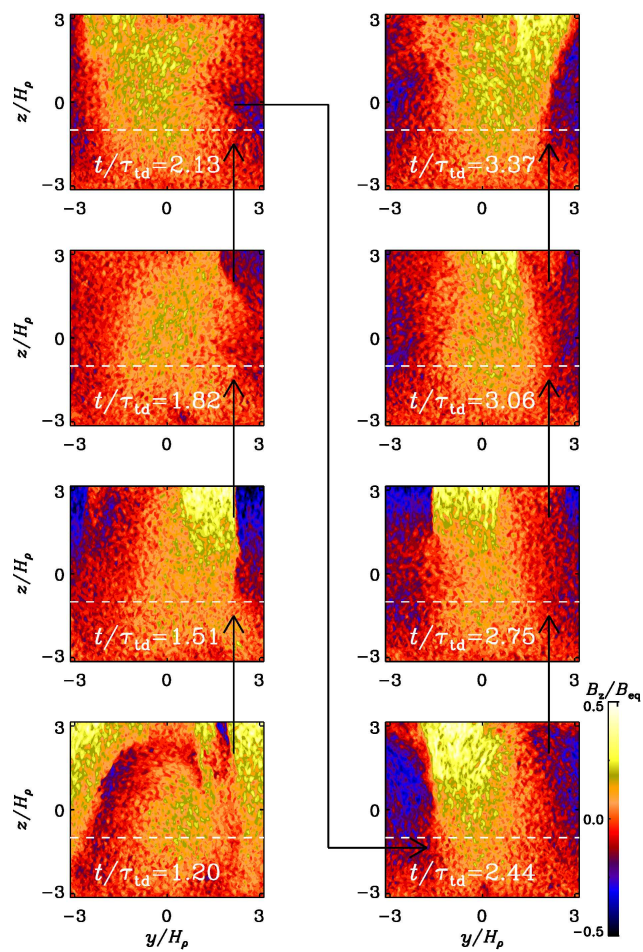}
\end{center}\caption[]{
Time evolution of $B_{z}/\Beq$ in the $yz$ plane
through $x/H_\rho=\pi$ for Run~RM1.
}\label{pbz_yz}\end{figure}

The magnetic surface structures are formed by a redistribution
of magnetic flux so that regions of highly concentrated magnetic field
are separated by regions of low magnetic field. This effect can be seen
in \Fig{pbz_yz} where we show a visualization of
$B_z/\Beq$ in the same temporal and spatial frame as \Fig{pbx_yz}.

\subsection{Relation to downflows}

\begin{figure}\begin{center}
\includegraphics[width=.8\columnwidth]{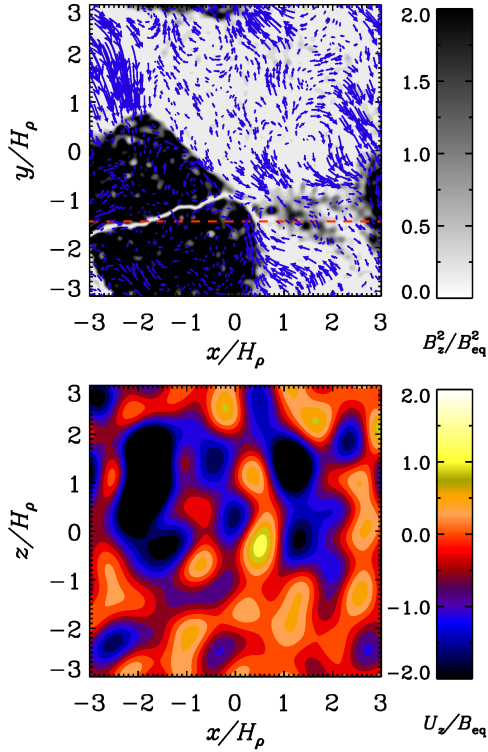}
\end{center}\caption[]{
Upper panel: velocity vectors, $\bra{U_x}_{k_{6}}$ and $\bra{U_y}_{k_{6}}$ (blue arrows) plotted
on a gray-scale representation of $B_{z}^2/\Beq^2(z)$ at the surface and $t/\tautd=2.46$
for Run~RM1.
Lower panel: visualization of $\bra{U_z}_{k_{6}}$ on $xz$ plane at $y/H_\rho=-1.5$
(red dashed line on the upper panel).
}\label{pbz_one}\end{figure}

As follows from previous related studies of
forced turbulence \citep{Mit14}, the magnetic flux
concentrations tend to form in regions with downflows.
To study this effect we plot in the upper panel of \Fig{pbz_one}
the large-scale horizontal velocity,
$\bra{(U_x,U_y)}_{k_{6}}$ (blue arrows), together with a gray-scale
representation of $B_{z}^2/\Beq^2(z)$ at the surface for $t/\tautd=2.46$.
Here, $\bra{\cdot}_{k_{6}}$ denotes Fourier filtering, applied to obtain
smoother contours.
We see that there are positions where the horizontal velocity
around or near the edge of the each spot is high.
Furthermore, the horizontal velocity is small where the field is strong,
which is consistent with the presence of downflows.
This is shown in the lower panel of \Fig{pbz_one}, where we plot
$\bra{U_z}_{k_{6}}$ in an $xz$ plane through $y/H_\rho=-1.5$.
There are indeed clear downflows below the magnetic flux concentration.
This is in agreement with previous studies of magnetic field concentrations
in one-layer turbulence.

\subsection{Scale separation}

\begin{figure}\begin{center}
\includegraphics[width=.99\columnwidth]{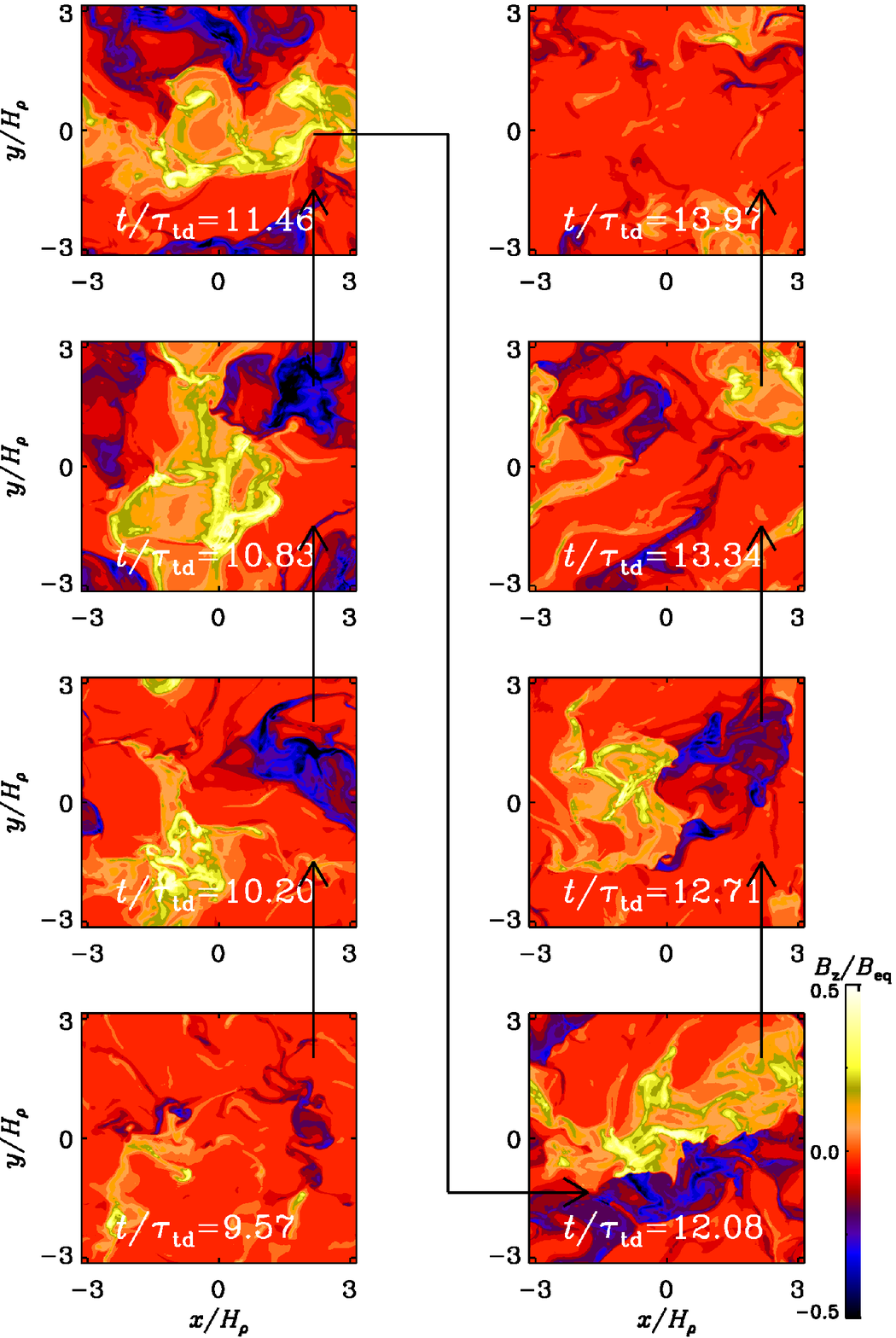}
\end{center}\caption[]{
Same as \Fig{pbz_xy} but for Run~RM1k with $k_{f}=5$.
}\label{pbz_xy_k}\end{figure}

Previous studies of magnetic flux concentrations in turbulence with weak
imposed magnetic field have shown that NEMPI forms magnetic concentrations
only when the scale separation ratio, $\kf/k_{1}$, is about 15 or larger
\citep{BGJKR14}.
Therefore, we perform a simulation with $\kf/k_1=5$ to study whether
the formation of structures is still possible in such a model.
\FFig{pbz_xy_k} presents the visualization of $B_{z}$
at the surface of the box for such a simulation.
Surprisingly, in our stratified two-layer forcing model,
the bipolar magnetic structures continue to form, and
follow a similar evolution as in the case with higher scale separation ratio.
The only difference is the time delay in the formation of
the first structure and their irregular and fast motions.

The other interesting case is when we apply a forcing profile that
is helical in the entire domain (Run~RM1zs).
In such a system, we expect the formation of a bipolar structure
at much earlier times, because the large-scale dynamo is now allowed to work
in the entire domain so we should observe propagating dynamo waves
at all depths.
Our results confirm this already at a time of around $0.37 \tautd$, when
a magnetic structure develops at the surface and the evolution of the structures
occurs faster than in the two-layer simulations.

\subsection{Evolution of bipolar structures}

To investigate the evolution of the bipolar magnetic structures in more detail,
we study the motion in the vicinity of the magnetic structures.
In the early stage of the formation of bipolar structures,
they tend to have round yin-yang shapes
and each polarity rotates clockwise, independently of each other.
When the structures move close enough to each other, their motion
is no longer independent.
After the formation of the elongated structures (see \Fig{pbz_xy} at
$t/\tautd=1.51$ and $1.82$), one can see an
anti-clockwise rotation at the border between opposite polarities
(at $t/\tautd=2.13$ and $2.44$ in this figure), which
tends to break the connection and destroys the elongated structure.
We suggest that the clockwise rotation of structures is due to the presence
of a strong large-scale magnetic helicity associated with the structure.
In other words, the traveling dynamo waves reach the surface
and affect the evolution and motion of the magnetic structures.
The anti-clockwise rotation, however, might be driven either by some
instability, which occurs when opposite magnetic fields come close
to each other (e.g., an instability similar to tearing instability
during reconnection) or it might be caused by
the interaction of two rotating polarities that are now coupled.

\begin{figure*}\begin{center}
\includegraphics[width=.51\columnwidth]{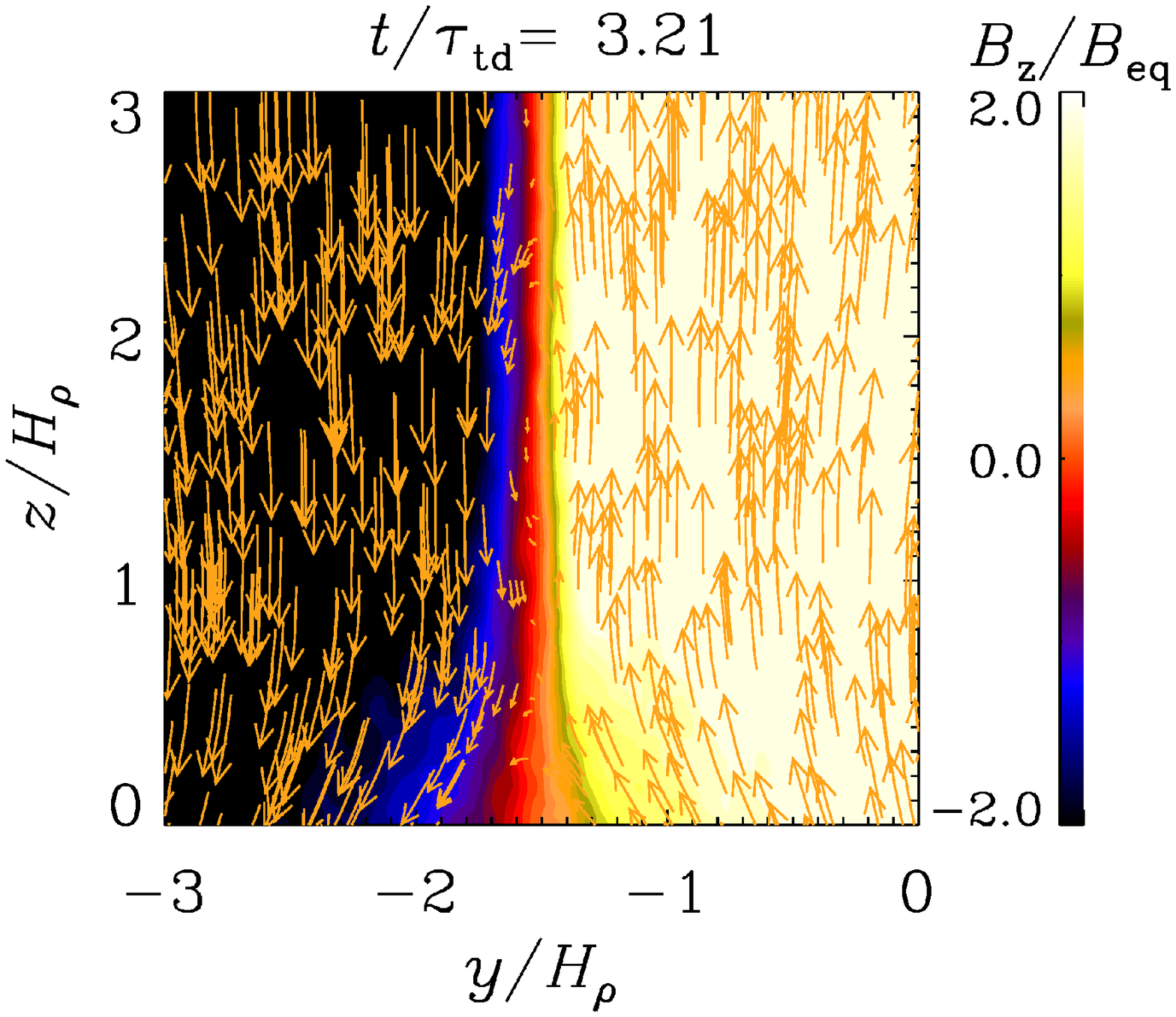}
\includegraphics[width=.51\columnwidth]{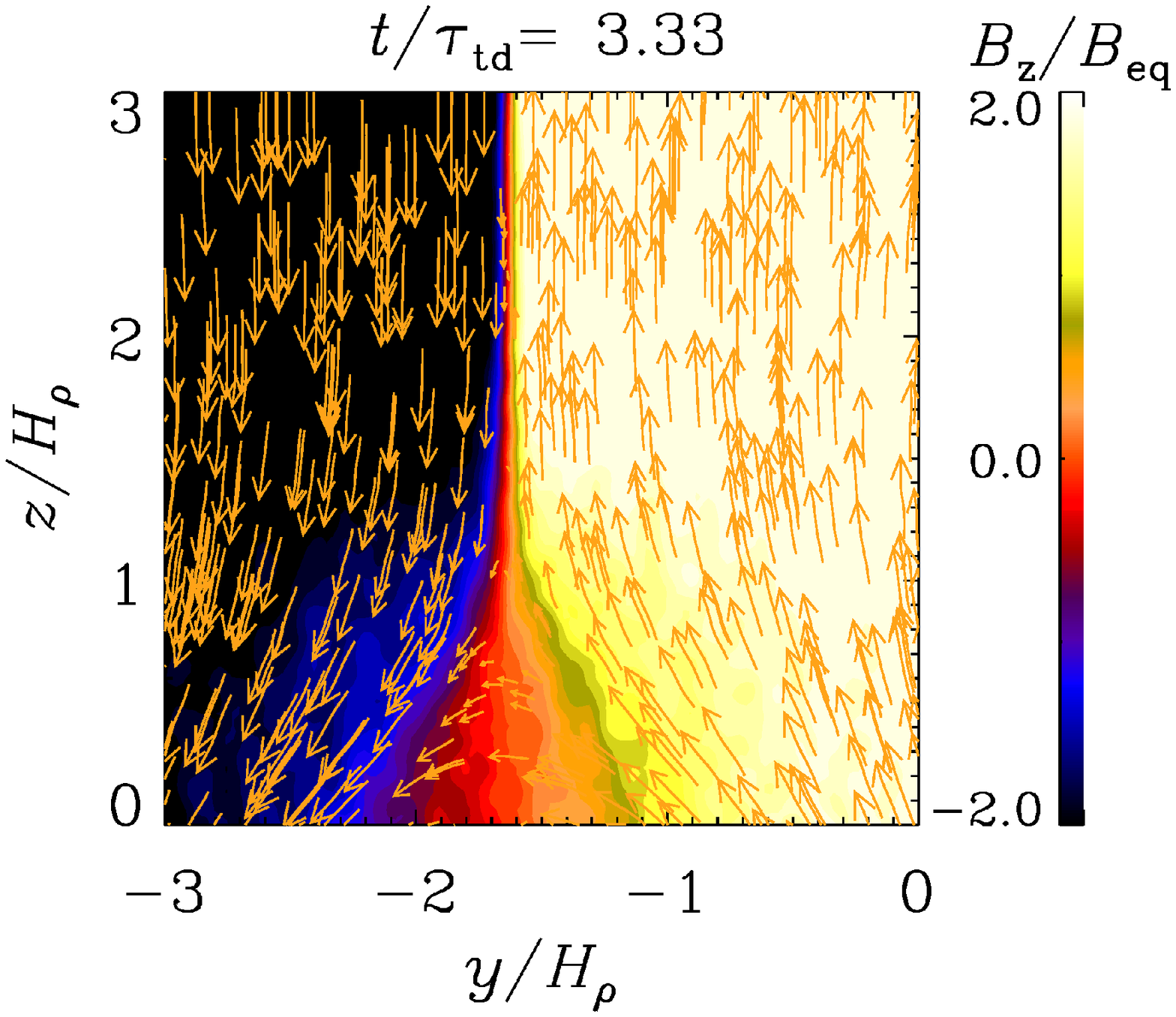}
\includegraphics[width=.51\columnwidth]{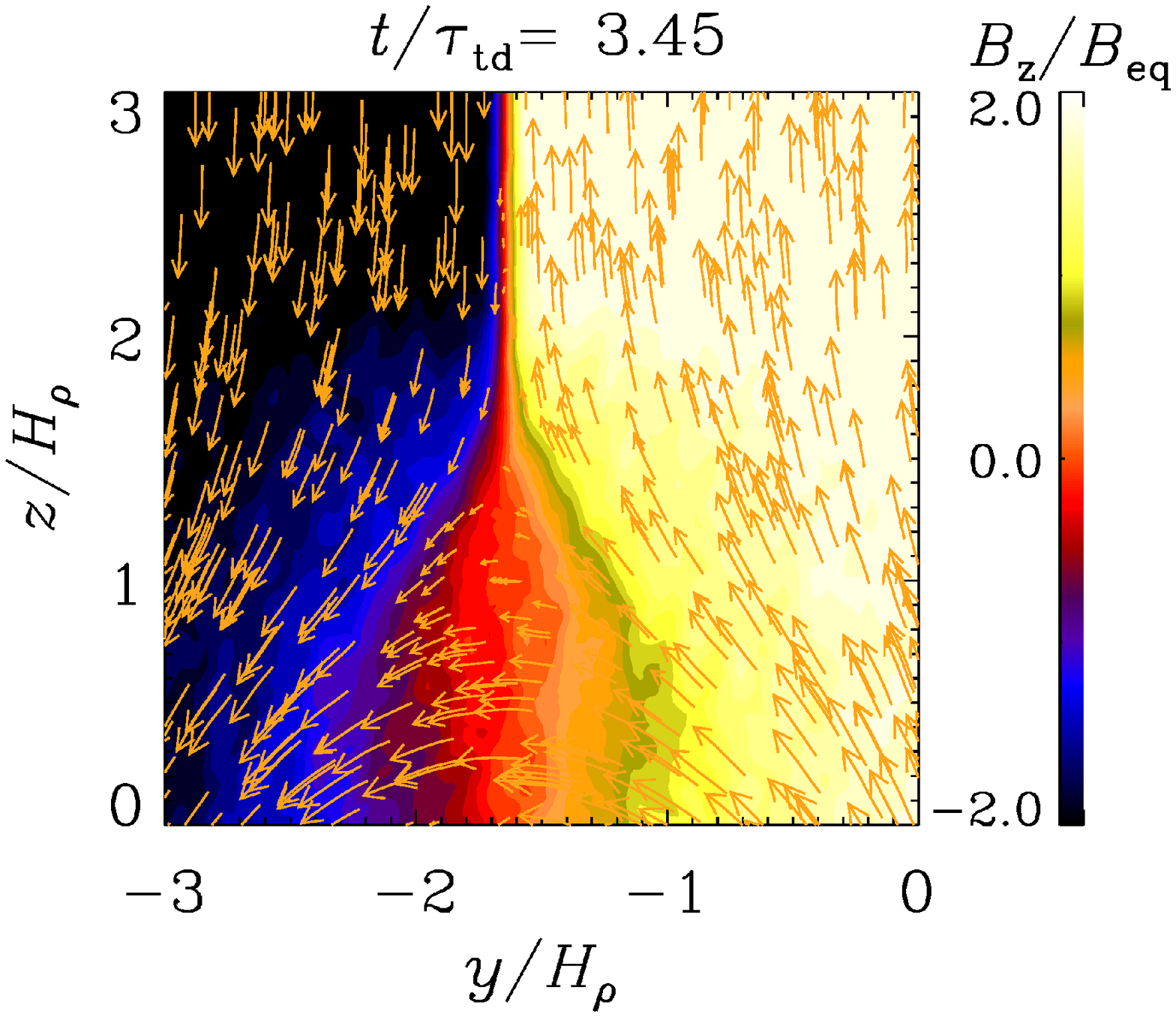}
\includegraphics[width=.51\columnwidth]{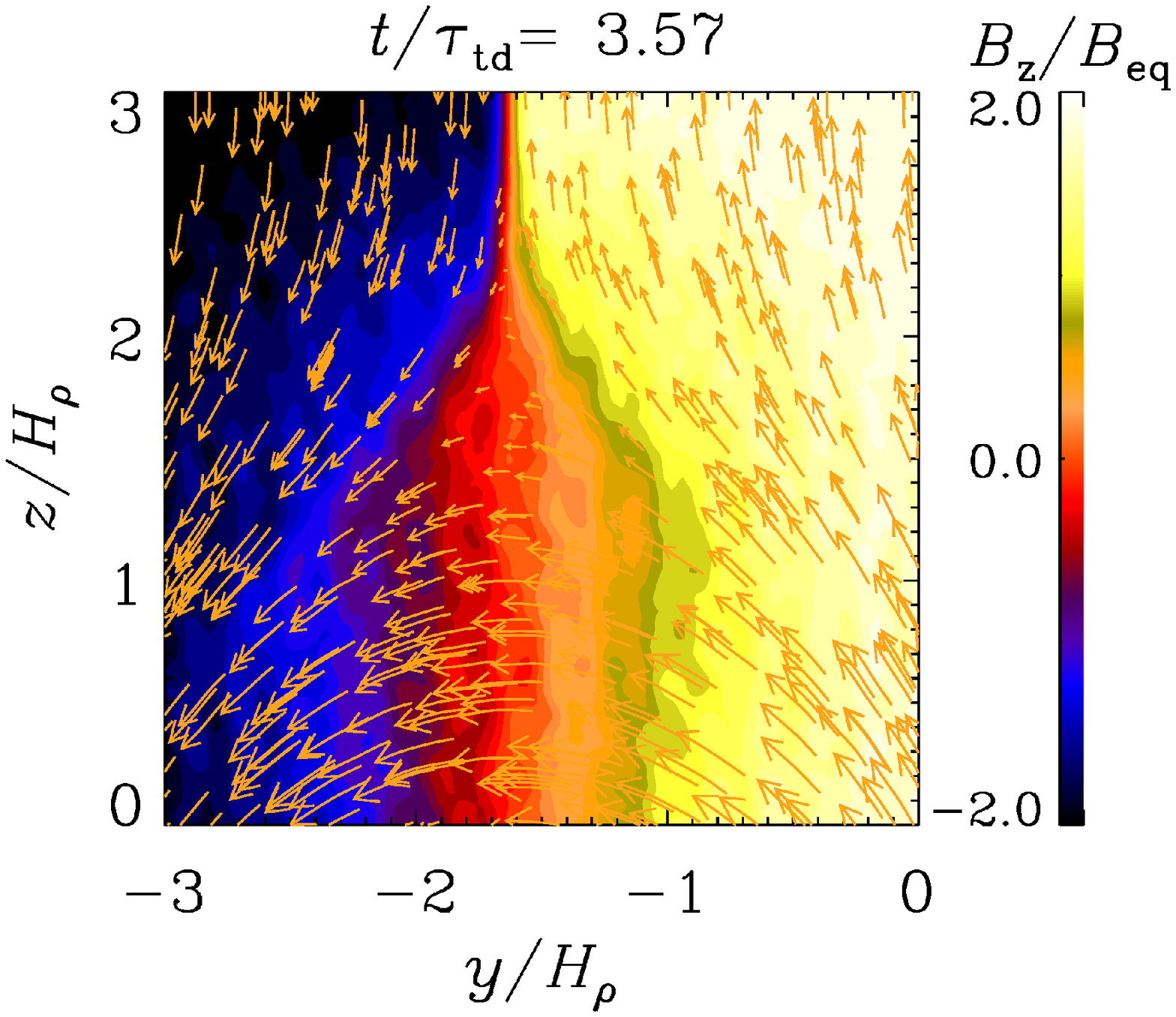}
\end{center}\caption[]{
Time evolution of $\mean{B_{z}}/\Beq$, together with $\mean{B_{y}}/\Beq$
and $\mean{B_{z}}/\Beq$ vectors for Run~RM1.
}\label{pvarm}\end{figure*}

\subsection{The effect of rotation}

In this work we also study the effect of rotation on the formation
and the evolution of the magnetic structures.
We perform simulations with different Coriolis numbers,
$\Co=0.03$, $0.3$, $0.7$, and $1.4$.
Table~\ref{Tab3} shows the parameters of these runs.
Previous studies showed that for $\Co$ larger than 0.1, rotation
suppresses the formation of magnetic structures by NEMPI,
see \cite{Los13} and \cite{Jabbari14}.
However, our present study shows that in our two-layer dynamo model,
magnetic structures survive for $\Co$ as large as 1.4.
It should also be noted that the combination of
stratification and rotation leads to the generation of
an additional contribution to the kinetic helicity in the system
\citep{KR80,KR03,Jabbari14}.
This contribution is either constructive if $\Co<0$
(producing extra positive helicity) or destructive if $\Co>0$
(producing negative helicity).
This could modify the dynamo action, but in the present case
the Coriolis number is still too small for the rotation-induced
helicity to be important; see Fig.~5 of \cite{Jabbari14}.

One of the possible reasons for the existence of magnetic structures for
moderate rotation rates ($\Co \leq 1.4$) is the large-scale dynamo that
increases the magnetic flux.
Indeed, NEMPI cannot create a new flux,
but can only redistribute it by forming magnetic concentrations in
certain small regions.
Since the dynamo systematically produces new magnetic flux,
and NEMPI redistributes it,
the magnetic concentrations survive even for a moderate rotation.

\begin{table}\caption{
Summary of the runs with rotation.
}\vspace{12pt}\centerline{\begin{tabular}{lllcccccl}
Run &$\Co$&$\theta$& $\lambda/\etatz k_1^2$ \\
\hline
{\bf RM1}& {\bf 0} &{\bf 0}&\;\;0.0122  \\
R1       &   0.037 &    0  &0.041  \\
R2       &   0.37  &    0  &0.040  \\
R3       &   0.74  &    0  &0.033 \\
R4       &   0.37  &$\pi/4$&0.040  \\
R5       &   0.37  &$\pi/2$&0.040  \\
R6       &$\!\!\!\!\!-0.37$ &    0  &0.040  \\
R7       &   1.4   &    0  &0.015 \\
%
%
\label{Tab3}\end{tabular}}
\end{table}

It turns out that in the presence of rotation with $\Co>0$, there is a delay
in the formation of bipolar structures and the development of their shape during
the early stage.
In the case without rotation, the structure has spherical-like shape
in the early stage of formation before it becomes elongated.
This does not happen at finite rotation with $\Co>0$.
In the presence of rotation, even in the early stage of
bipolar structure formation, it has a random elongated shape, which changes
rapidly in time.
In the presence of rotation
it takes more time for the magnetic structures
to become intense and concentrated.

\section{Reconnection in the upper layers}

\begin{figure}\begin{center}
\includegraphics[width=.65\columnwidth]{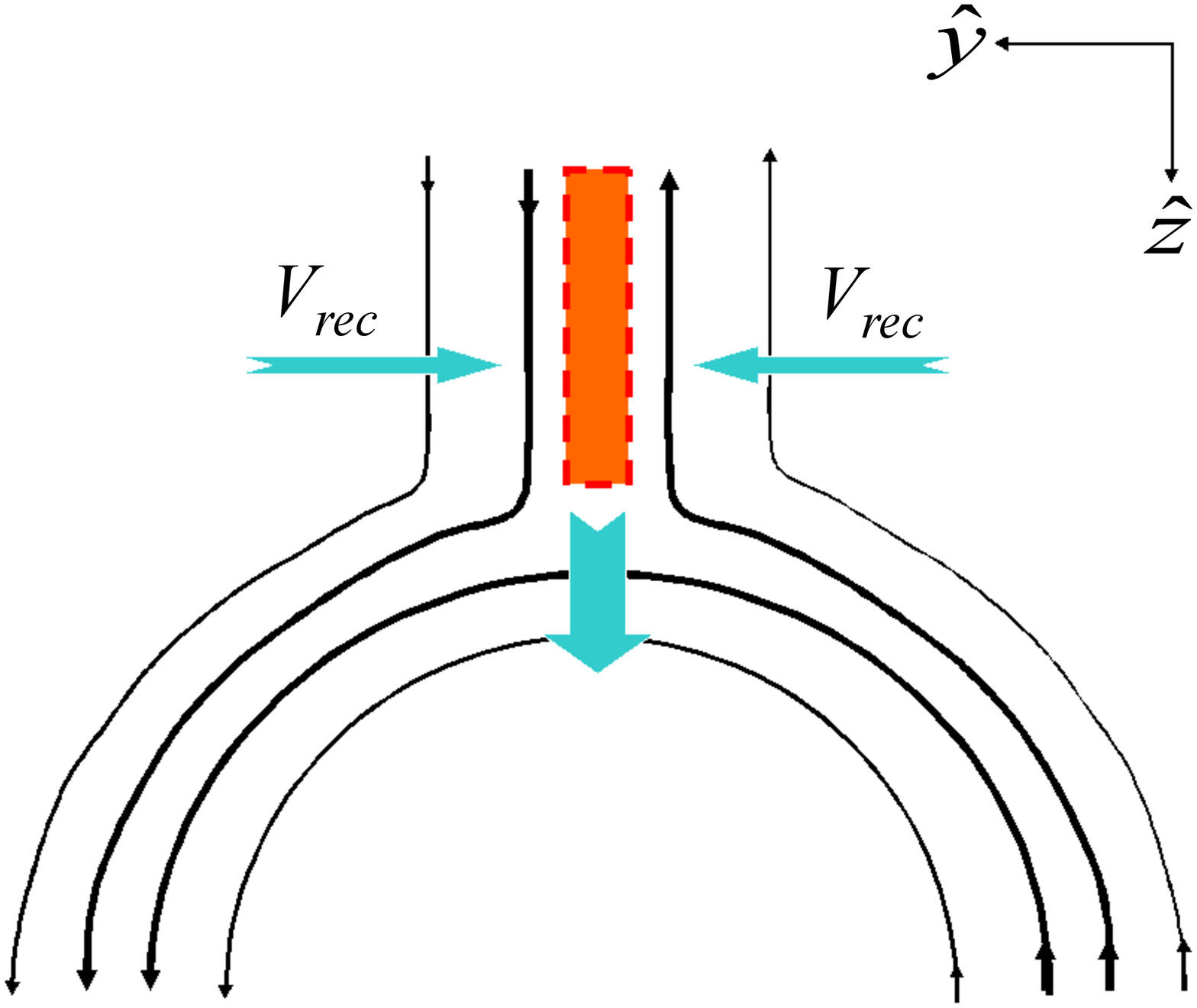}
\includegraphics[width=.6\columnwidth]{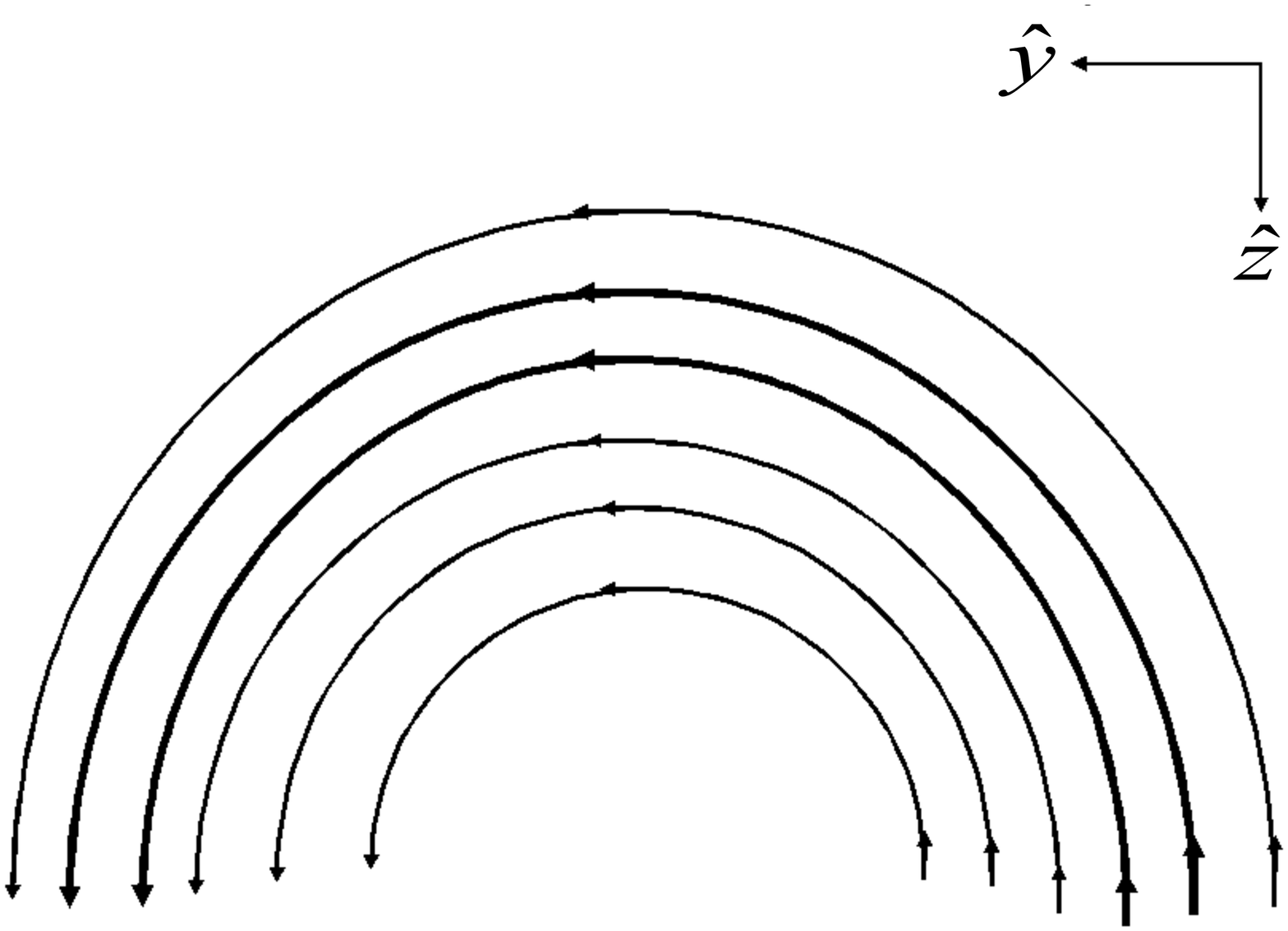}
\end{center}\caption[]{
Upper panel: formation of a current sheet before the reconnection.
Lower panel: magnetic configuration after the reconnection.
}\label{rec}\end{figure}

The production of sharp fronts can be seen in \Figs{pbz_xy}{pbz_yz},
where we plot $\mean{B_{z}}/\Beq$ in two different planes.
During the evolution of the magnetic structures, the bipolar regions
evolve into stripes of opposite polarities separated by a current sheet;
see \Fig{bz3D}.
In \Fig{pvarm} we zoom in on the sharp front in
$\mean{B_{z}}/\Beq$, where we also see vectors of $\mean{B_{y}}/\Beq$
and $\mean{B_{z}}/\Beq$ in the $yz$ plane for our reference run.
By comparing the field lines with \Fig{rec}, one can see
a similar reconfiguration of magnetic field lines during spot evolution.
It is clear from this figure that field lines with opposite signs of
$\mean{B_{z}}/\Beq$ are reconnected and a $y$ component of the magnetic
field is generated.

On a timescale that lies somewhere between the resistive diffusion
and Alfv\'en timescales, magnetic reconnection occurs which causes the
magnetic field topology to change and leads to the conversion of magnetic
energy to thermal energy, kinetic energy and even particle acceleration;
see the sketch in \Fig{rec}.
According to Sweet-Parker theory \citep{P57,S58,S69,P94}, hereafter SP theory,
the rate of reconnection, $V_{\rm rec}$ depends on the Lundquist number, $S$:
\begin{equation}
V_{\rm rec} = V_{\rm A} S^{-1/2}.
\end{equation}

In the turbulent regime of reconnection, $V_{\rm rec}$ is independent
of the Ohmic resistivity \citep{LV99}; see also \cite{ELV11}.
This conclusion is supported by numerical simulations \citep{KLV09}.
According to \cite{LV99}, the upper limit for the reconnection rate is:
\begin{equation}
V_{\rm rec} \sim V_{\rm A} M_{\rm A}^{2},
\end{equation}
where $M_{\rm A}=u_{\rm rms}/V_{\rm A}$ is the Alfv\'en Mach number.

For large Lundquist numbers, and in the turbulent regime of reconnection,
$V_{\rm rec}$ is independent of $S$.
This conclusion is confirmed by recent
numerical simulations \citep{LUS09,HB10}.
For $S>10^4$, the Sweet-Parker current
sheet is unstable \citep{B86,LCD05,LSS12,Oishi15}.
For spontaneous magnetic reconnection, according to magnetohydrodynamic
numerical simulations \citep{LUS09,HB10,B13}, the rate of the
reconnection, $V_{\rm rec}$, for $S>10^4$ is of the order of
\begin{equation}
V_{\rm rec} \sim (1 - 3) \times 10^{-2} V_{\rm A} .
\end{equation}

To determine $V_{\rm rec}$, one can use two approaches.
In one approach, the value of the inflow in the vicinity
of the current sheet is measured (as sketched in \Fig{rec}).
For a turbulent plasma, on the other hand, one can use a more general
and accurate method.
In this approach, one uses Ohm's law:
\begin{equation}
\eta \mu_0\JJ = \EE + \UU \times \BB ,
\end{equation}
so that the rate of the reconnection, $V_{\rm rec}$, can be determined as
$V_{\rm rec} \simeq V_E$, where
\begin{equation}
V_E = {\langle | \EE | \rangle \over \langle | \BB | \rangle}
= {\eta \langle | \mu_0\JJ| \rangle - \langle | \UU \times \BB |
\rangle \over \langle | \BB | \rangle},
\label{recE}
\end{equation}
and angular brackets denote averaging along the $z$ direction
(along the largest side of the current sheet,
i.e., perpendicular to the current, see \Fig{rec}).
Thus, the method of the determining $V_{\rm rec}$
in numerical simulations is as follows:
(i) find the region with the current sheet that is separating
magnetic fields of opposite polarities;
(ii) use different instants of
the formation of the current sheet and determine the value of $V_{\rm rec}$,
the length of the current layer, ${\cal L}_{z}$, in the $z$ direction,
the Alfv\'en speed $V_{\rm A}$ in these instants.
Finally, we determine $S^{-1/2}$ and $M_{\rm A}^2=u_{\rm rms}^2/V_{\rm A}^2$,
and compare these quantities with the obtained value of $V_{\rm rec}/V_{\rm A}$.
We recall that the length of the current sheet ${\cal L}_{z}$
enters in the definition $S=V_{\rm A} {\cal L}_z /\eta$, and
the time when ${\cal L}_{z}$ reaches its maximum value
marks the starting time of reconnection.

To determine $V_{\rm rec}$ we use $x$-averaged data,
average over the interval $(z_1,z_2)$, where ${\cal L}_{z}=z_2-z_1$
is the length of the current sheet. Next, we measure the value of $V_{\rm rec}$
as $V_E(y^*)$ and $V_{\rm in}(y^*)$, while $y^*$ is a point, which is
in the vicinity of the current sheet.
\FFig{rec}, upper panel, already shows the position
of $V_{\rm rec}$, which is the same for both $V_{\rm in}$ and $V_E$.

\begin{table*}\caption{
Summary of the reconnection parameters.
}\vspace{4pt}\centerline{\begin{tabular}{lrlccccccccccccccl}
Run & $\Rm$ & $\eta$ &$t/\tautd$ & ${\cal L}_{z}$ & $u_{\rm rms}/c_{\rm s}$ & $V_E/c_{\rm s}$ & $V_{\rm A}/c_{\rm s}$ &${\rm M}_{\rm A}$  & $S$ & $S^{-1/2}$ & $V_E/V_{\rm A}$ &${\rm M}_{\rm A}^2$ & $V_{\rm in}/c_{\rm s}$  & $V_{\rm in}/V_{\rm A}$\\
\hline
RM0& 10 & $3 \times 10^{-4}$ &0.97& 0.025 & 0.125& 0.04 & 0.319 & 0.39 & 26 &0.196 & 0.127& 0.149&0.033 &0.104\\ 
D1 & 16 & $2 \times 10^{-4}$ &0.65& 0.0246& 0.12 & 0.0136& 0.231 &0.52& 28 &0.188& 0.0589& 0.27  & 0.0101&0.0437\\ 
%
%
%
D1 & 16 & $2 \times 10^{-4}$ &0.84& 0.0247& 0.118& 0.0206& 0.299 &0.4 & 37 &0.165& 0.0689& 0.156 & 0.0212&0.0709\\ 
%
%
D1 & 16 & $2 \times 10^{-4}$ &0.97& 0.123 & 0.104& 0.0067& 0.169 &0.62& 104&0.098& 0.0397& 0.379 & 0.0075&0.0444\\ 
%
RM1 & 50  & $5.7 \times 10^{-5}$ &2.87& 0.0493&  0.115 & 0.0107 & 0.228 & 0.5 & 187 & 0.073 & 0.047  & 0.254& 0.0154 &0.068 \\ 
%
%
RM1 & 50  & $5.7 \times 10^{-5}$ &2.92& 0.0493&  0.120 & 0.0151 & 0.275 & 0.44& 226 & 0.067 & 0.055  & 0.191& 0.0211 &0.077\\ 
%
%
RM1 & 50  & $5.7 \times 10^{-5}$ &2.99& 0.0986&  0.122 & 0.0099 & 0.323 & 0.38& 531 & 0.043 & 0.031  & 0.143& 0.0176  &0.055 \\ 
%
%
RM1 & 50  & $5.7 \times 10^{-5}$ &3.04& 0.1478&  0.126 & 0.0120 & 0.395 & 0.32& 973 & 0.032 & 0.03 & 0.102& 0.0184 &0.047 \\ 
%
%
RM1 & 50  & $5.7 \times 10^{-5}$ &3.09& 0.1971&  0.123 & 0.0108 & 0.441 & 0.28& 1449& 0.026 & 0.025 & 0.078& 0.0145 &0.033 \\ 
%
%
RM1 & 50  & $5.7 \times 10^{-5}$ &3.16& 0.1971&  0.117 & 0.0071 & 0.435 & 0.27& 1429& 0.027 & 0.016 & 0.072& 0.0093 &0.022 \\ 
%
%
RM1 & 50  & $5.7 \times 10^{-5}$ &3.21& 0.5914&  0.118 & 0.0121 & 0.395 & 0.3 & 3893& 0.016 & 0.031 & 0.089& 0.0137 &0.035 \\ 
%
%
RM1 & 50  & $5.7 \times 10^{-5}$ &3.34& 1.7987&  0.109 & 0.0061 & 0.322 & 0.34& 9653& 0.010 & 0.019 & 0.115& 0.0112 &0.035 \\ 
%
%
RM1 & 50  & $5.7 \times 10^{-5}$ &3.46& 1.3306&  0.103 & 0.0060 & 0.239 & 0.43& 5300& 0.014 & 0.025 & 0.186& 0.0092 &0.039 \\ 
%
%
RM1 & 50  & $5.7 \times 10^{-5}$ &3.58& 0.9363&  0.096 & 0.0075 & 0.176 & 0.55& 2747& 0.019 & 0.043 & 0.298& 0.0104 &0.059 \\ 
%
%
RM1 & 50  & $5.7 \times 10^{-5}$ &3.71& 0.4435&  0.092 & 0.0053 & 0.130 & 0.71& 961 & 0.032 & 0.041 & 0.501& 0.0043 &0.033 \\ 
%
%
RM1 & 50  & $5.7 \times 10^{-5}$ &3.83& 0.3450&  0.094 & 0.0101 & 0.089 & 1.06& 512 & 0.044 & 0.11 & 1.116& 0.0107 &0.12 \\ 
%
RM2 & 130  &$2 \times 10^{-5}$  &1.6 & 2.4147& 0.092 & 0.0039 & 0.183 & 0.50 &22095& 0.0067 & 0.0213 & 0.253 & 0.0058& 0.032 \\ 
RM2 & 130  &$2 \times 10^{-5}$  &1.65& 1.7987& 0.093 & 0.0089 & 0.181 & 0.51 &16278& 0.0078 & 0.0492 & 0.264 & 0.0094& 0.052 \\ 
RM2 & 130  &$2 \times 10^{-5}$  &1.7 & 1.4045& 0.094 & 0.0074 & 0.187 & 0.50 &13132& 0.0087 & 0.0396 & 0.253 & 0.0091& 0.049 \\ 
RM2 & 130  &$2 \times 10^{-5}$  &1.75& 1.1581& 0.091 & 0.0080 & 0.194 & 0.47 &11234& 0.0094 & 0.0412 & 0.220 & 0.0094& 0.049 \\ 
RM2 & 130  &$2 \times 10^{-5}$  &1.8 & 0.8131& 0.092 & 0.0080 & 0.187 & 0.49 &7603 & 0.0115 & 0.0428 & 0.242 & 0.0094& 0.050 \\ 
%
%
RM2 & 130  &$2 \times 10^{-5}$  &1.9 & 0.5174& 0.088 & 0.0051 & 0.151 & 0.58 &3906 & 0.0160 & 0.0338 & 0.340 & 0.0074& 0.049 \\ 
%
%
%
RM3 & 260  & $1. \times 10^{-5}$ &3.04& 0.468& 0.09 & 0.005& 0.198&0.455&9266 &0.011& 0.024& 0.207& 0.005&0.024 \\ 
%
%
RM3 & 260  & $1. \times 10^{-5}$ &3.24& 0.665& 0.088& 0.006& 0.178&0.494&11837&0.009& 0.033& 0.245& 0.005&0.028 \\ 
RM3 & 260  & $1. \times 10^{-5}$ &3.33& 2.538& 0.086& 0.004& 0.165&0.521&41877&0.005& 0.026& 0.272& 0.004&0.024 \\ 
RM3 & 260  & $1. \times 10^{-5}$ &3.43& 1.848& 0.086& 0.005& 0.159&0.541&29383&0.006& 0.032& 0.293& 0.003&0.021 \\ 
RM3 & 260  & $1. \times 10^{-5}$ &3.53& 1.010& 0.086& 0.006& 0.154&0.558&15554&0.008& 0.035& 0.312& 0.003&0.021 \\ 
RM3 & 260  & $1. \times 10^{-5}$ &3.63& 0.542& 0.087& 0.022& 0.094&0.926&5094 &0.014& 0.233& 0.857& 0.007&0.070 \\ 
%
%
%
%
\label{Tab2}\end{tabular}}
\end{table*}

\begin{figure}\begin{center}
\includegraphics[width=.99\columnwidth]{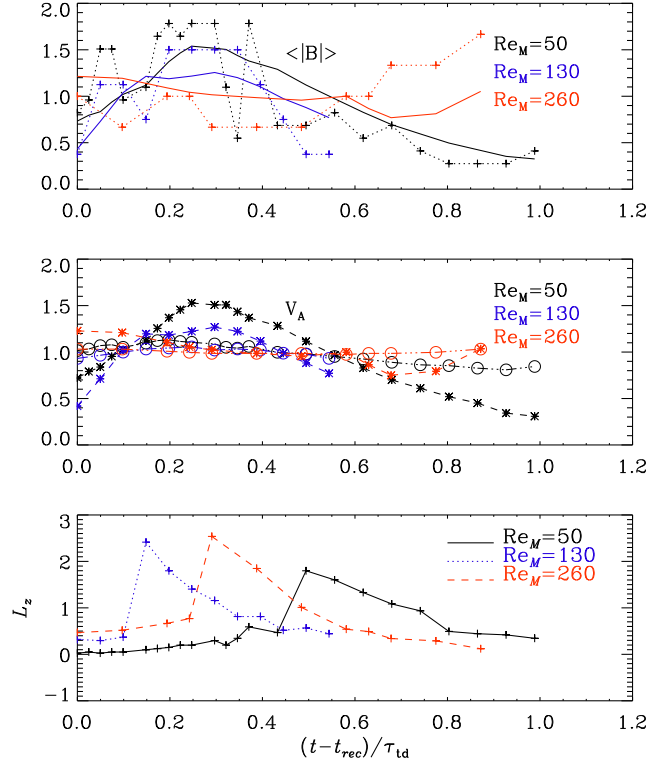}
\end{center}\caption[]{
Upper panel:
time evolutions of $\langle | \BB | \rangle$ (solid) and
$\langle | \EE | \rangle$ (dotted),
both normalized by their time-averaged value.
Different colors are related to three different $\Rm$ (Runs~RM1, RM2, and RM3).
Middle panel:
time evolutions of $V_{\rm A}/\langle V_{\rm A} \rangle$ (stars),
and $u_{\rm rms}/\langle \urms \rangle$ (circles) for Runs~RM1 (black), RM2 (blue), and
RM3 (red).
Lower panel:
time evolutions of
${\cal L}_{z}k_{1}$ (triangles) for Runs~RM1 (black), RM2 (blue), and
RM3 (red).
 }\label{pt}\end{figure}

The resulting values of $V_{\rm rec}$ are summarized in \Tab{Tab2}.
For comparison, we measure both the velocity $V_{\rm E}$
using \Eq{recE} and the incoming velocity $V_{\rm in}$
in the vicinity of the current sheet (in the case of a two-dimensional
flow in the $xz$-plane it is in the $y$ direction).
We normalize these velocities by $\langle V_{\rm A} \rangle$,
because $V_{\rm A}$ varies with time (see \Fig{pt}).
In this section, in order to be able to compare the results of
different runs, we use $(t-t_{\rm rec})/\tautd$ for the normalized time, where
$t_{\rm rec}$ is the time when the reconnection starts for each individual Run.
\FFig{pt} presents the time evolution of the different quantities
in three separate panels: $\langle| \BB |\rangle$ and
$\langle | \EE | \rangle$ in the upper panel,
$V_{\rm A}/\langle V_{\rm A} \rangle$ and
$u_{\rm rms}/\langle \urms \rangle$ in the middle panel, and finally
${\cal L}_{z}k_{1}$ in the lower panel.
The different colors represent  different values of $\Rm$.
One can see that, $\urms/\langle \urms \rangle$ does not change strongly with time
and $V_{\rm A}/\langle V_{\rm A} \rangle$ changes when $\Rm$ is smaller
($\Rm=50$).
This implies that the major change in $S$ comes from the change in
the length of the current sheet (see the lower panel of \Fig{pt}).

\begin{figure}\begin{center}
\includegraphics[width=.99\columnwidth]{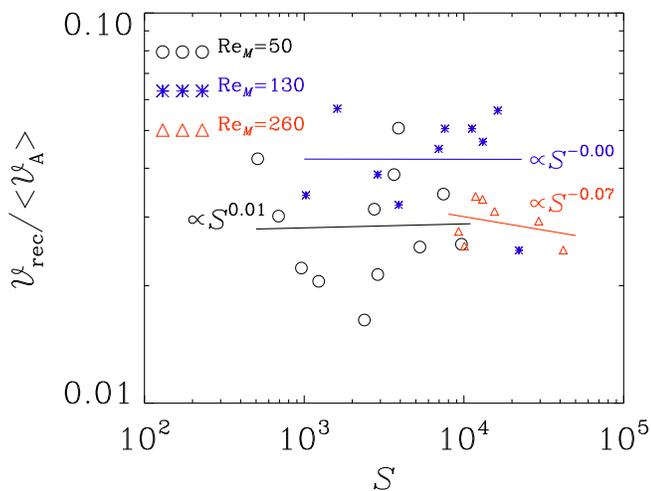}
\end{center}\caption[]{
Reconnection rate $V_{\rm E} / \langle V_{\rm A} \rangle$
normalized by the mean Alfv\'en speed versus $S$.
The colors represent the value of $\Rm$ (Runs~RM1 (circles), RM2 (stars),
and RM3 (triangles)).
}\label{vrecvas_fit_2p}
\end{figure}

\begin{figure}\begin{center}
\includegraphics[width=.98\columnwidth]{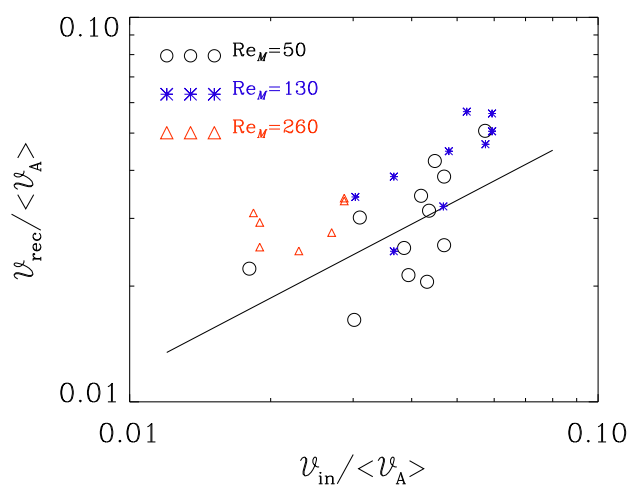}
\end{center}\caption[]{
Reconnection rate $V_{\rm E} / \langle V_{\rm A} \rangle$
normalized by the mean Alfv\'en speed versus $V_{\rm in} / \langle V_{\rm A} \rangle$.
The colors represent the value of $\Rm$ (Runs~RM1 (circles), RM2 (stars),
and RM3 (triangles)). 
 }\label{vrecvin}\end{figure}

To check which regime of magnetic reconnection is appropriate,
we plot in \Fig{vrecvas_fit_2p} $V_{\rm E}$ as a function of $S$.
By comparing the curves for different magnetic Reynolds numbers,
$\Rm=50$, $130$, $260$, it is clear that our data points are consistent with
the turbulent regime of reconnection \citep{LUS09,HB10,LSS12,B13},
where the reconnection rate is nearly independent of $S$.
\Fig{vrecvin} demonstrates that the reconnection rates obtained from
the measurements of $V_{\rm in}$ and $V_{\rm E}$ are similar.

\begin{figure}\begin{center}
\includegraphics[width=.98\columnwidth]{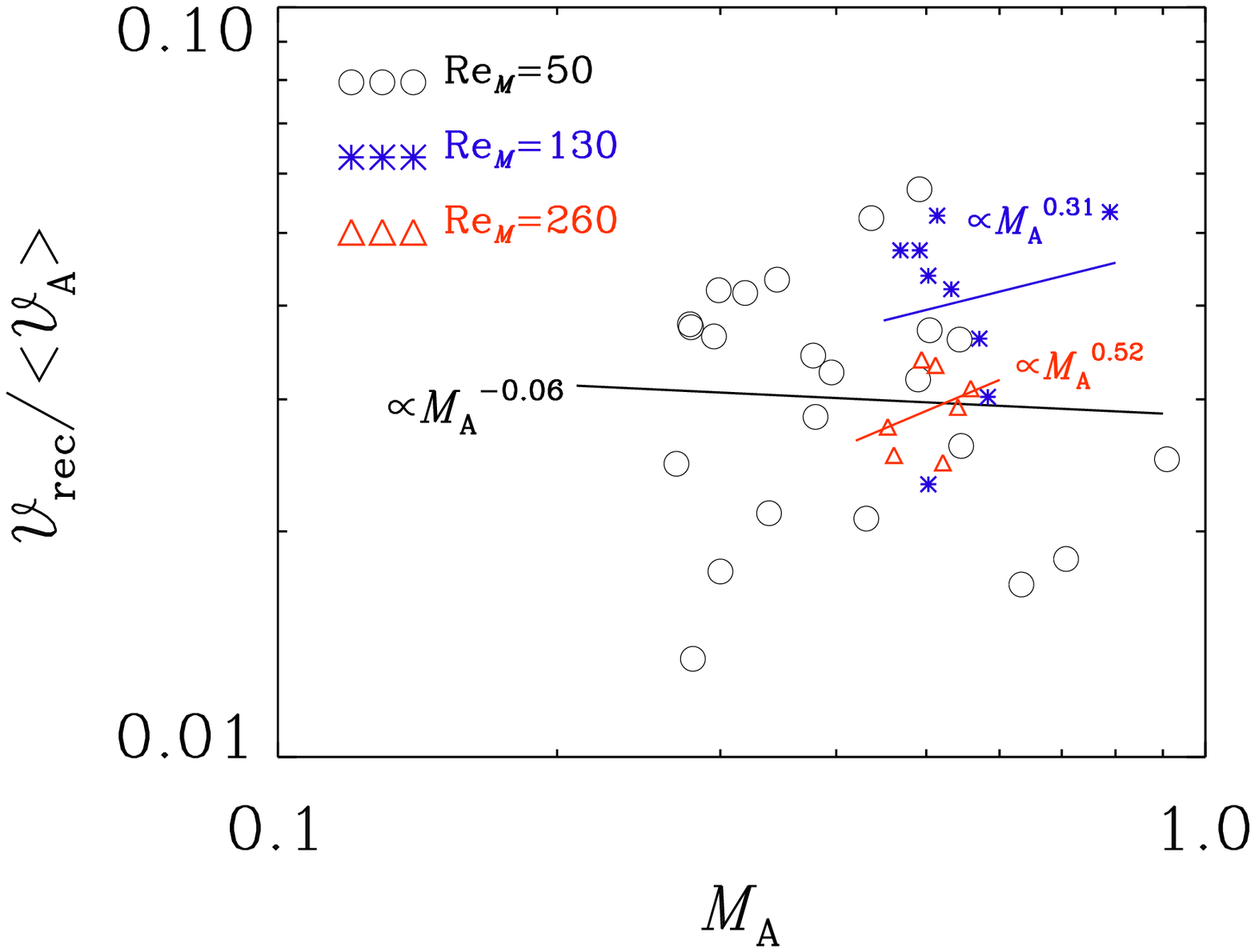}
\end{center}\caption[]{
$V_{\rm E}$
normalized by $\langle V_{\rm A} \rangle$ as a function of $\MA$.
The solid line presents the best linear fit.
Different colors present different values of $\Rm$
($\Rm$=50 (Run~RM1), black circles, $\Rm$=130 (Run~RM2), blue stars,
and $\Rm$=260 (Run~RM3), red triangles).
}\label{vrecma}\end{figure}

The dependence of the reconnection rate on $\MA$ is shown in \Fig{vrecma}.
To compare the resulting data from our simulations with the model
of \cite{LV99}, we also plot the linear fit to the data points
in \Fig{vrecma}.
It is clear that our data points
strongly deviate from the predicted $\MA^2$ line,
and are thus inconsistent with \cite{LV99}.
However, $V_{\rm rec}$ is weakly dependent on
the Ohmic resistivity (see \Figs{vrecvas_fit_2p}{vrecma}), in
agreement with \cite{LV99}.

\section{Conclusions}

In the present paper, we have extended the results of \cite{Mit14} to
higher magnetic Reynolds numbers and have investigated the effects
of rotation at different Coriolis numbers.
Our results demonstrate that in the two-layer model
with helical forcing in the lower part and non-helical forcing in the upper,
sharp bipolar spots form at the surface, expand and then
develop stripy structures.
The observed effects are similar for different values of $\Rm$ and $\Co$.
In our present simulations, for $\Co$ as large as 1.4, we still observe
the formation of the intense bipolar structures.
This value is significantly larger than what was previously obtained
in studies of magnetic flux concentrations in rotating
turbulence with an imposed weak magnetic field.
One of the plausible explanations for this is the large-scale dynamo
which increases magnetic flux.
By contrast, NEMPI cannot produce new flux, but only redistribute it,
forming magnetic concentrations in small regions.
Thus, the large-scale dynamo systematically
produces new magnetic flux, while NEMPI redistributes it.
This could be the reason why the magnetic concentrations survive even
for moderate rotation.
Although there are dynamo-generated magnetic fields in our simulations, which
is more realistic compared to several previous models with
an imposed magnetic field, we still
observe evidence for downflows at the locations
of magnetic structure formation,
similar to \cite{BGJKR14}, \cite{Mit14}, and \cite{Jabbari15}.

What is surprising is the long lifetime of the resulting bipolar regions,
which exceeds several turbulent diffusion times.
We suggest that the main reason why these intense magnetic structures survive
longer is the magnetic reconnection phenomenon in the vicinity
of the current sheet between opposite magnetic polarities.
We have determined the reconnection rate for a range of different parameters
and have shown that for high Lundquist numbers, $S>10^3$,
the measured reconnection rate is nearly independent of $S$.
This result is
consistent with recent numerical simulations in a turbulent regime of
reconnection performed by other groups \citep{LUS09,HB10,LSS12,B13}.
The measured reconnection rate is weakly dependent on the Alfv\'en Mach
number, $M_{\rm A}^2$, which is inconsistent with predictions of \cite{LV99}.
On the other hand, the reconnection rate is also weakly dependent on
the Ohmic resistivity, in agreement with \cite{LV99}.

In the present work, we have also investigated the effects of
varying the scale separation ratio, $\kf/k_{1}$.
Contrary to earlier studies of NEMPI, for $\kf/k_{1}$ as small as
five, bipolar magnetic structures still form.
Our previous studies of an unipolar magnetic concentrations with an imposed
weak mean magnetic field have shown that,
although the effective magnetic pressure (the sum of turbulent and non-turbulent
contributions) becomes negative even for moderate scale separation ratio (about
3--5), the large-scale instability (NEMPI) is excited only if the scale separation
ratio is large enough ($>15$).
This suggests that the phenomenon we find in our DNS cannot be understood solely
in terms of NEMPI.

In the more complicated two-layer system with a dynamo-generated magnetic
field, two instabilities (dynamo and probably NEMPI or the magnetic buoyancy
instability) may be excited.
We stress that in both, our two-layer model with dynamo-generated magnetic field
and in turbulent systems with an imposed magnetic field where NEMPI
is known to be excited, strong density stratification plays an important role.
Furthermore, there is evidence for
the existence of downflows at the locations of magnetic structures
in both systems.
However, in our two-layer model the formation of bipolar regions is
still observed for smaller scale separation ratios than what was required
for a turbulent system with an imposed magnetic field where NEMPI was excited.

The process maintaining the bipolar structures may be
related to or associated with NEMPI.
It may also be possible that the positive magnetic pressure associated with
the strong dynamo-generated magnetic field in nonlinear stage of evolution
is responsible for the formation of
the sharp interface of the bipolar structures found in the upper layers,
where the plasma beta is no longer very large.
However, a conclusive answer cannot be given at present.
To arrive at a more definitive conclusion regarding the mechanism of
bipolar structure formation, it would be desirable to measure
the effective magnetic pressure tensors in our two-layer model
and to study its parameter dependency in more detail.
This requires the development of an adequate test-field method.
This is a subject of a separate study.

\section*{Acknowledgements}
We appreciate valuable advise by Alexander Schekochihin and Nuno Loureiro
to study magnetic reconnection in our two-layer system.
We also acknowledge valuable discussions with Andrey Beresnyak and
Alexandre Lazarian as well as constructive remarks from the referee.
This work was supported in part by
the Swedish Research Council Grants No.\ 621-2011-5076 (AB,SJ),
2012-5797 (AB),  and 638-2013-9243 (DM), as well as
the Research Council of Norway under the FRINATEK grant 231444
(AB, NK, IR).
We acknowledge the allocation of computing resources provided by the
Swedish National Allocations Committee at the Center for
Parallel Computers at the Royal Institute of Technology in
Stockholm, the High Performance Computing Center North in Ume\aa,
and the Nordic High Performance Computing Center in Reykjavik.

\newcommand{\yastroph}[2]{ #1, astro-ph/#2}
\newcommand{\ycsf}[3]{ #1, {Chaos, Solitons \& Fractals,} {#2}, #3}
\newcommand{\yepl}[3]{ #1, {Europhys.\ Lett.,} {#2}, #3}
\newcommand{\yaj}[3]{ #1, {AJ,} {#2}, #3}
\newcommand{\yjgr}[3]{ #1, {J.\ Geophys.\ Res.,} {#2}, #3}
\newcommand{\ysol}[3]{ #1, {Sol.\ Phys.,} {#2}, #3}
\newcommand{\yapj}[3]{ #1, {ApJ,} {#2}, #3}
\newcommand{\ypasp}[3]{ #1, {PASP,} {#2}, #3}
\newcommand{\yapjl}[3]{ #1, {ApJ,} {#2}, #3}
\newcommand{\yapjs}[3]{ #1, {ApJS,} {#2}, #3}
\newcommand{\yija}[3]{ #1, {Int.\ J.\ Astrobiol.,} {#2}, #3}
\newcommand{\yan}[3]{ #1, {Astron.\ Nachr.,} {#2}, #3}
\newcommand{\yzfa}[3]{ #1, {Z.\ f.\ Ap.,} {#2}, #3}
\newcommand{\ymhdn}[3]{ #1, {Magnetohydrodyn.} {#2}, #3}
\newcommand{\yana}[3]{ #1, {A\&A,} {#2}, #3}
\newcommand{\yanas}[3]{ #1, {A\&AS,} {#2}, #3}
\newcommand{\yanar}[3]{ #1, {A\&A Rev.,} {#2}, #3}
\newcommand{\yass}[3]{ #1, {Ap\&SS,} {#2}, #3}
\newcommand{\ygafd}[3]{ #1, {Geophys.\ Astrophys.\ Fluid Dyn.,} {#2}, #3}
\newcommand{\ygrl}[3]{ #1, {Geophys.\ Res.\ Lett.,} {#2}, #3}
\newcommand{\ypasj}[3]{ #1, {Publ.\ Astron.\ Soc.\ Japan,} {#2}, #3}
\newcommand{\yjfm}[3]{ #1, {J.\ Fluid Mech.,} {#2}, #3}
\newcommand{\ypepi}[3]{ #1, {Phys.\ Earth Planet.\ Int.,} {#2}, #3}
\newcommand{\ypf}[3]{ #1, {Phys.\ Fluids,} {#2}, #3}
\newcommand{\ypfb}[3]{ #1, {Phys.\ Fluids B,} {#2}, #3}
\newcommand{\ypp}[3]{ #1, {Phys.\ Plasmas,} {#2}, #3}
\newcommand{\ysov}[3]{ #1, {Sov.\ Astron.,} {#2}, #3}
\newcommand{\ysovl}[3]{ #1, {Sov.\ Astron.\ Lett.,} {#2}, #3}
\newcommand{\yjetp}[3]{ #1, {Sov.\ Phys.\ JETP,} {#2}, #3}
\newcommand{\yphy}[3]{ #1, {Physica,} {#2}, #3}
\newcommand{\yaraa}[3]{ #1, {ARA\&A,} {#2}, #3}
\newcommand{\yanf}[3]{ #1, {Ann. Rev. Fluid Mech.,} {#2}, #3}
\newcommand{\yrpp}[3]{ #1, {Rep.\ Prog.\ Phys.,} {#2}, #3}
\newcommand{\yprs}[3]{ #1, {Proc.\ Roy.\ Soc.\ Lond.,} {#2}, #3}
\newcommand{\yprt}[3]{ #1, {Phys.\ Rep.,} {#2}, #3}
\newcommand{\yprl}[3]{ #1, {Phys.\ Rev.\ Lett.,} {#2}, #3}
\newcommand{\yphl}[3]{ #1, {Phys.\ Lett.,} {#2}, #3}
\newcommand{\yptrs}[3]{ #1, {Phil.\ Trans.\ Roy.\ Soc.,} {#2}, #3}
\newcommand{\ymn}[3]{ #1, {MNRAS,} {#2}, #3}
\newcommand{\ynat}[3]{ #1, {Nature,} {#2}, #3}
\newcommand{\yptrsa}[3]{ #1, {Phil. Trans. Roy. Soc. London A,} {#2}, #3}
\newcommand{\ysci}[3]{ #1, {Science,} {#2}, #3}
\newcommand{\ysph}[3]{ #1, {Solar Phys.,} {#2}, #3}
\newcommand{\ypr}[3]{ #1, {Phys.\ Rev.,} {#2}, #3}
\newcommand{\ypre}[3]{ #1, {Phys.\ Rev.\ E,} {#2}, #3}
\newcommand{\ypnas}[3]{ #1, {Proc.\ Nat.\ Acad.\ Sci.,} {#2}, #3}
\newcommand{\yicarus}[3]{ #1, {Icarus,} {#2}, #3}
\newcommand{\yspd}[3]{ #1, {Sov.\ Phys.\ Dokl.,} {#2}, #3}
\newcommand{\yjcp}[3]{ #1, {J.\ Comput.\ Phys.,} {#2}, #3}
\newcommand{\yjour}[4]{ #1, {#2}, {#3}, #4}
\newcommand{\yprep}[2]{ #1, {\sf #2}}
\newcommand{\ybook}[3]{ #1, {#2} (#3)}
\newcommand{\yproc}[5]{ #1, in {#3}, ed.\ #4 (#5), #2}
\newcommand{\pproc}[4]{ #1, in {#2}, ed.\ #3 (#4), (in press)}
\newcommand{\pprocc}[5]{ #1, in {#2}, ed.\ #3 (#4, #5)}
\newcommand{\pmn}[1]{ #1, {MNRAS}, to be published}
\newcommand{\pana}[1]{ #1, {A\&A}, to be published}
\newcommand{\papj}[1]{ #1, {ApJ}, to be published}
\newcommand{\ppapj}[3]{ #1, {ApJ}, {#2}, to be published in the #3 issue}
\newcommand{\sprl}[1]{ #1, {PRL}, submitted}
\newcommand{\sapj}[1]{ #1, {ApJ}, submitted}
\newcommand{\sana}[1]{ #1, {A\&A}, submitted}
\newcommand{\smn}[1]{ #1, {MNRAS}, submitted}



\label{lastpage}
\end{document}